\documentstyle[preprint,revtex]{aps}

\begin{document}

\font\fortssbx=cmssbx10 scaled \magstep2
\hskip.5in \raise.1in\hbox{\fortssbx University of Wisconsin - Madison}
\hfill$\vcenter{\hbox{\bf MAD/PH/740}
                \hbox{February 1993}}$ }

\vspace{.25in}

\begin{title}
\large \bf Empirical Determination of the Very High Energy\\
 Heavy Quark Cross Section from Non-Accelerator Data
\end{title}
\author{\large M.C.~Gonzalez-Garcia, F.~Halzen and R.A.~V\'azquez}
\begin{instit}
Department of Physics,University of Wisconsin, Madison,  WI 53706, USA
\end{instit}
\author{\large E.~Zas}
\begin{instit}
Departamento de F\'\i sica de Part\'\i culas,
Universidad de Santiago\\  E-15706 Santiago de Compostela, Spain
\end{instit}

\thispagestyle{empty}

\begin{abstract}
To cosmic rays incident near the horizon the Earth's atmosphere represents a
beam dump with a slant depth reaching 36000~g~cm$^{-2}$ at $90^\circ$. The
prompt decay of a heavy quark produced by very high energy cosmic ray showers
will leave an unmistakable signature in this dump. We translate the failure of
experiments to detect such a signal into an upper limit on the heavy quark
hadroproduction cross section in the energy region beyond existing
accelerators. Our results disfavor any rapid growth of the cross section or the
gluon structure function beyond very conservative estimates based on
perturbative QCD.
\end{abstract}


\newpage

Understanding the production of heavy quarks, especially the relatively light
charm quark, is a subject that is at the forefront of particle physics for
several reasons. Leptonic decays of the charm and $b$-quark is the source of
high momentum electrons and muons which represent critical backgrounds in the
search for new phenomena at present and future hadron colliders. Perspectives
for very interesting neutrino physics at future colliders, including the
direct observation of tau neutrinos, depend critically on the production of
charm\cite{deruj}. Because of its intermediate mass the charm quark provides a
bridge between light and heavy quark physics. Precisely for this reason the
mechanisms for charm production are believed not to be well understood. A
perturbative calculation lies beyond the scope of existing QCD technology
because it requires resummation of large logarithms of $1/x$ with
$x\simeq m_c/\sqrt{s}$\cite{collins}. Even the leading order perturbative
calculations, which require inclusion of $O(\alpha_s^2)$ and $O(\alpha_s^3)$
diagrams, are unreliable because the results are sensitive to the assumed
quark mass and the renormalization scale. Because charm is at high energies
predominantly produced by gluons, the cross section also critically depends on
the low-$x$ behavior of the gluon structure function which is poorly known or
undetermined depending on the energy\cite{nason89}. Charm production is a
great laboratory to study these issues on the interphase of perturbative and
non-perturbative QCD. The experimental status of open charm production is
unfortunately rather unsettled. Large uncertainties are especially associated
with the production of charm particles in the Feynman-$x$, $x_F \rightarrow 1$
region which is poorly covered by the highest energy colliders.

Cosmic ray particles have been observed up to energies exceeding
$10^{20}$~eV\cite{Akeno}, about 100 times the SSC energy, and thus have a
unique potential for studying the cross-section for charmed particle
production at energies impossible to achieve with current technology in
controlled experiments. More importantly, the leptonic decay of charm
particles into high energy muons produces a gold-plated signature leading to
the rather unusual circumstance of doing a ``clean'' particle physics
experiment with cosmic particles. Cosmic rays with energy in excess of
$100$~TeV initiate air showers which can be studied with sea-level particle
detectors. The detected flux is a steeply falling function of zenith angle
because the depth of atmosphere traversed by a cascade reaching the ground
rises rapidly from 1030 to 36000~g~cm$^{-2}$ as the zenith angle varies from
zero to 90 degrees. Thus near the horizon ($90^\circ$) close to a thousand
radiation lengths of matter separate the interaction from the detector and the
configuration of the experiment is identical to that of any accelerator-based
beam dump experiment. Most secondaries such as pions and kaons are absorbed in
the dump and only penetrating particles, such as muons and neutrinos produced
in the initial interaction, reach the detector. For very high energy
interactions the decay of charm particles is the dominant source of high
energy secondary muons. So, counting high energy muons at large zenith angles,
typically larger than 70 degrees,  determines the charm cross section. There
is no background from the semi-leptonic decay of pions and kaons which, as a
result of time dilation, interact and lose energy rather than decay into high
energy muons. This background has been extensively studied
and is well understood\cite{DEIS}.

It should also be pointed out that the muon signature can be sharply defined,
even when using a conventional air shower array as a detector. The high energy
muon will traverse the atmosphere and occasionally lose energy by catastrophic
photon bremsstrahlung. If the photon shower is produced close to the detector
it will be recorded and is referred to as a horizontal air shower. The origin
of such a signal can be verified as it must be i) independent of zenith angle
unlike any potential background, ii) it must have a low muon content as the
shower is purely electromagnetic, and iii) as it must consist of shower
detected in the vicinity of the shower maximum. All features can be
experimentally verified and incorporated in the triggers selecting horizontal
air showers. Thus the measurement of high energy muon fluxes can undoubtedly
provide conclusive data on the production of charm.

It is the purpose of this letter to illustrate how several existing
installations can collect data relevant to charm production. Most importantly,
the Akeno air shower detector in Japan has published in 1985 an upper limit on
the flux of muon-poor horizontal air showers with energy in excess of
100~TeV\cite{Akenobound}. The experiment has since tripled its statistics and
has not found a signal. It is straightforward to translate the upper limit
into an upper limit on the charm cross section. The translation requires
knowledge of the primary cosmic ray spectrum which is well measured in the
energy region under consideration, the published effective area and exposure
time of the detector, the energy loss of high energy muons and the
well-understood structure of the electromagnetic shower radiated by the muon.
Details are given in reference\cite{munmad}. Our results are shown in
Figs.~\ref{oursigma}-\ref{modelsigma}.

Figure~\ref{oursigma} shows our final results as upper bounds on the charm
cross section obtained by comparing a variety of extrapolations of accelerator
data \cite{expdata} with the published and preliminary bounds on horizontal
air shower fluxes. We fitted the accelerator data in Fig.~\ref{oursigma} with
a function that reproduces the charm cross section up to some energy $E_c$,
$O(10^2)\le E_c \le O(10^3)$~GeV. Above this energy we extrapolated the cross
section using a variety of asymptotic parametrizations. The transition energy
and the asymptotic behaviour were varied as independent parameters. The
fastest increase of the charm cross section consistent with existing
accelerator data and with the negative results of the full Akeno search
behaves roughly as $\log^2(s)$. We verified that the result is insensitive to
the parametrization. The derivation requires an assumption for the
$x_F$-dependence of the inclusive charm cross section. The bands representing
the bounds in Figs.~\ref{oursigma} and \ref{modelsigma} reflect this
ambiguity, which has been modelled using the $x_F$-distributions for
$D$-mesons from extreme predictions ranging from the quark-gluon string model
to perturbative QCD. In the end the
ambiguity is only of the order of a factor of 2. It does mean however that
cross sections for models with unusually flat $x_F$-dependence can possibly
violate this bound. Conversely, deriving the bound with a flatter
$x_F$-distribution will strengthen it. Such an effect could be associated with
the forward production of $\Lambda_c$. The forward baryon is more effective at
producing high energy muons and as a consequence the absence of a positive
signal can only be accommodated with a smaller production cross section. This
trend is however offset by a reduction of the semi-leptonic branching ratio by
a factor of three.

It is customary in cosmic ray experiments to quote the directly accessible
number of electrons and positrons $N_e$ in the horizontal shower rather than
the energy of the photon. To a good approximation the energy of the photon, in
GeV-units, is given by 2~$N_e$. The primary energy of interacting cosmic
particles sampled by a detector with threshold $N_e = 10^5$--$10^6$ is
determined by the energy dependence of the initial flux, the Feynman-$x$
distribution of the charmed particles, and their semi-leptonic decay
distribution. A simple analytic estimate leads to the expectation that the
data is sensitive to initial proton energies one order of magnitude larger
that the muon energy\cite{multip}. Our detailed simulation reproduces this
result and we have therefore established a bound on the charm cross section to
energies of order $E_{\rm lab}\approx 10^3$--$10^4$~TeV $(\sqrt{s}\approx
1$--5~TeV). The prediction for the number of horizontal air showers
corresponding to the bounds on the charm cross section shown in
Fig.~\ref{oursigma}, is shown in Fig.~\ref{showers} along with present
experimental upper limits on the horizontal air shower rates. These limits are
established on the basis of less than one detected shower with 1986 and
present statistics for $N_e>10^5$. Also shown is the limit corresponding to
less than one detected shower with $N_e>10^6$ with present statistics.


{}From a theoretical point of view our results are rather surprising. As shown
in Fig.~\ref{modelsigma} the preliminary bound based on the present Akeno
statistics is reproduced by a straightforward perturbative calculation of the
charm cross section to order O$(\alpha_s^3)$ using the KMRSD0 structure
functions\cite{KMRS} which correspond to a conservative estimate of the number
of low-$x$ gluons. The bound does not allow for large enhancements in the
cross section or a rapid growth of the gluon structure function at small
values of $x$ beyond what is implied by the scaling $1/x$ prediction assumed
in the KMRSD0 parametrization. Both had been widely predicted in the
literature. The fact that the next to leading order contribution to the cross
section is of the same order as the leading order result led to speculation of
further enhancements associated with the resummation of large logarithms
$\log(\sqrt s/m_c)$. This point is made in a different way in
Fig.~\ref{modelsigma} where we illustrate that predictions in the literature
tend to violate the bound we obtained. For illustration we have plotted the
prediction for the charm cross section obtained from perturbative QCD to order
$O(\alpha_s^3)$ with structure functions growing as $1/x$ (KMRSD0), and
$1/x^{1.5}$ (KMRSD- and Nason\cite{LHCnason}) together with the non
perturbative result from the quark-gluon string model for two sets of
parameters ($\alpha_{\Psi}=0,-2.18$)\cite{QGSM}.

Is it possible to relax this bound? In deriving our bound we also assumed that
all high energy cosmic rays are protons. It is experimentally known that the
cosmic ray spectrum is dominated by protons up 100~TeV. Above this energy the
composition is unknown. Introducing heavy primaries in the spectrum has, to a
first approximation, the effect of replacing protons of energy $E$ by $A$
protons of energy $E/A$. These are less efficient at producing muons above the
fixed threshold set by the experiment and our bound on the production cross
section will be relaxed. We illustrate the sensitivity to composition
introducing an extreme assumption. In order to accommodate cosmic ray
observations without having to claim the appearance of "anomalies" such as the
celebrated Centauro phenomena, the Fuji \cite{Yuda} group has assumed that
40\% of the cosmic rays are heavy nuclei. We will follow this lead and make
the further extreme assumption that all nuclei are iron. The net effect is to
relax the bound in Fig.~\ref{oursigma} by a factor 1.6-2 depending, on the
energy, allowing also a faster growth.

In summary, we do not know of any credible way to weaken the bound we
presented by more than a factor of 2 or so. The production of forward
particles will, on the contrary, considerably strengthen it. Evidence for
forward production of charmed baryons and mesons has been presented in several
experiments. Especially a component of forward $D$-mesons with a harder
energy spectrum should significantly strengthen the upper limit on charm
hadroproduction we derived from horizontal air showers.

Consequences of relevance to future experiments both in collider physics and
high energy astrophysics can be derived from our result. Our bound translates
into a charm cross section of the order $\sigma\approx 2-3$ mb at LHC-SSC
energies. As pointed out the perspectives for the direct observation of tau
neutrinos, predominantly produced via the leptonic decay of strange-charmed
$D$-mesons, depend crucially on the value of this cross section\cite{deruj}.
A back of the envelope estimate yields a number of neutrino interactions of
the order 500(200) at LHC(SSC) per year .

Finally, the production of high energy neutrinos by charmed particles produced
in the atmosphere has been identified as a potential background in the high
energy neutrino telescopes presently under construction. The results derived
in this paper are directly relevant and reassuring. Also, existing
underground experiments are in principle sensitive to the high energy muons
discussed here \cite{munmad}.

\acknowledgments
We thank M. Nagano for his patience and generous help in explaining the Akeno
air shower array results and for communication of their preliminary results.
We also thank R. Fletcher, T. Stanev and T. Stelzer for enlightening
discussions and criticism. This work was supported by the University of
Wisconsin Research Committee with funds granted by the Wisconsin Alumni
Research Foundation, by the U.S.~Department of Energy under Contract
No.~DE-AC02-76ER00881, by the Texas National Research Laboratory Commission
under Grant No.~RGFY9273, by the ''Comit\'e Conjunto Hispano-Norteamericano''
and by the CICYT.

\figure{Total charm quark production cross section vs the CM and Lab energies.
The cross-hatched band represents the upper bound on the cross section obtained
by
requiring that extrapolations of the accelerator data accommodate the failure
of cosmic ray experiment to detect muon decay of charm particles by the Akeno
experiment using their full statistics. The dotted band represents the
corresponding bound for their published 1986 statistics.  The energy range
the Akeno array is sensitive to is shadowed. The width of the bands reflects
the use of a variety of $x_F$ distributions in  the derivation of the bounds.
The
upper (lower) edge of the bands correspond to  the steepest (flattest)
assumption for the $x_F$ distribution.  \label{oursigma}}

\figure{Integral number of horizontal showers (shower size $>N_e$) per second
and steradian as a function of shower size. The dashed and dotted areas
correspond to the minimum number of showers initiated by prompt muons from
charm decay for the cross sections shown in Fig.~\ref{oursigma}. The other
curves are the predictions for the different models in Fig.~\ref{modelsigma}
(using the same notation and value of the parameters). Also shown are the
limits from Akeno air shower experiment for $N_e=10^5$ for both 1986 statistics
and present full statistics, and the limit for $N_e=10^6$ with their present
statistics. \label{showers}}

\figure{Total charm quark production cross section vs the CM and Lab energies
for several models. Solid (dashed) line is the prediction from perturbative QCD
to order $O(\alpha_s^3)$ with KMRSD0 (KMRSD-) structure functions and $m_c=1.3$
GeV. The dotted line represents the result of the same calculation using the
structure functions from Ref.~\cite{LHCnason} and $m_c=1.6$ GeV . In all cases
the scale of both $\alpha_s$ and the structure functions is $Q^2=9$ GeV$^2$.
Dash-dotted lines are the non perturbative results from the quark-gluon string
model for two parameters values $\alpha_\psi=-2.2$ (upper) and $\alpha_\psi=0$
(lower).Also shown is a compilation of experimental  data from
Ref.~\cite{expdata} and our upper bounds (shaded bands). The bands span the
sensitivity range of the Akeno data. \label{modelsigma} }

\end{document}